\documentclass{PoS}
\title{Screening masses in the SU(3) pure gauge theory and universality}

\ShortTitle{Screening masses in the SU(3) pure gauge theory and universality}

\author{\speaker{R. Falcone$^a$}, R. Fiore$^a$, M. Gravina$^a$ and A. Papa$^a$\\ 
        \llap{$^a$}Dipartimento di Fisica, Universit\`a della Calabria,\\
        and Istituto Nazionale di Fisica Nucleare, Gruppo collegato di Cosenza\\
        I--87036 Arcavacata di Rende, Cosenza, Italy\\
        E-mail: \email{rfalcone,fiore,gravina,papa@cs.infn.it} }

\abstract{We determine from Polyakov loop correlators the screening masses in the deconfined 
phase of the (3+1)$d$ SU(3) pure gauge theory at finite temperature near the transition, 
for two different channels of angular momentum and parity. 
Their ratio is compared with that of the massive excitations with the same quantum numbers
in the 3$d$ 3-state Potts model in the broken phase near the transition point at zero 
magnetic field. 
Moreover we study the inverse decay length of the correlation between the real parts and 
between the imaginary parts of the Polyakov loop and compare the results with expectations
from perturbation theory and mean-field Polyakov loop models.}

\FullConference{The XXV International Symposium on Lattice Field Theory\\
		 July 30-4 August 2007\\
		 Regensburg, Germany}

\begin{document}

\section{Introduction}

In this work we compare the spectrum of the inverse decay lengths of Polyakov loop correlators 
in the (3+1)$d$ SU(3) gauge theory in the deconfined phase near the transition with the spectrum 
of massive excitations of the 3$d$ 3-state Potts model in the broken phase near the transition 
at zero magnetic field, which were determined in Ref.~\cite{FFGP06}.
The aim of the work is to verify if and to what extent the Svetitsky-Yaffe 
conjecture~\cite{Svetitsky:1982gs} also holds for theories which undergo a {\em weakly} first 
order phase transition, using mass ratios as a probe. In particular  we focus on the low-lying 
masses in two different sectors of parity and orbital angular momentum, 0$^+$ and 2$^+$.
We expect that, if universality would apply in strict sense, these spectra should exhibit 
the same pattern, as suggested by several numerical determinations in the 3$d$ 
Ising class~\cite{Caselle:1999tm,Caselle:2001im,Fiore:2002fj,Fiore-Papa-Provero-2003}. 

We extend our numerical analysis to temperatures far away from the transition
temperature $T_t$ in order to look for possible ``scaling'' of the fundamental masses
with temperature.
Moreover, we consider also the screening masses resulting from correlators of the real parts
and of the imaginary parts of the Polyakov loop. These determinations can represent useful
benchmarks for effective models of the high-temperature phase of SU(3), such as those based
on mean-field theories of the Polyakov loop, suggested by R.~Pisarski~\cite{Pisarski0101168}.

\section{Screening masses from Polyakov loop correlators}

Screening masses are defined as the inverse decay lengths of the Yukawa-like potential
between two static sources. They are generally determined through the
correlation of suitable operators. In our case correlations are between operators with 
different spatial separation.
The general large distance behavior for the correlation function 
$G(|z_1-z_2|)$, in an infinite lattice, is: 
\begin{equation}
G(|z_1-z_2|)= \sum_n a_n e^{-m_n |z_1-z_2|}\ ,
\label{corr-funct}
\end{equation}
where $m_0$ is the fundamental mass, while $m_1$, $m_2$, ... are higher masses
with the same angular momentum and parity quantum numbers of the fundamental mass.
On a periodic lattice the above equation must be modified by the inclusion of the so called
``echo'' term:
\begin{equation}
G(|z_1-z_2|)= \sum_n a_n\biggl[e^{-m_n |z_1-z_2|}+e^{-m_n (N_z-|z_1-z_2|)}\biggr]\ .
\label{corr-funct_echo}
\end{equation}   
The fundamental mass in a definite channel can be extracted from wall-wall 
correlators by looking for a plateau of the effective mass at large distances,
\begin{equation}
m_{\mbox{\footnotesize eff}}(z)= \ln \frac{G(z-1)}{G(z)} \ .
\label{m_eff}
\end{equation}

In the 0$^+$-channel, the connected wall-wall correlator in the $z$-direction is defined as
\begin{equation}
G(|z_1-z_2|)=\mbox{Re}\langle \bar{P}(z_1) \bar{P}(z_2)^\dagger \rangle -|\langle P \rangle|^2\ ,
\label{corr0}
\end{equation}
where
\begin{equation}
\bar{P}(z)=\frac{1}{N_xN_y}\sum_{n_x=1}^{N_x} \sum_{n_y=1}^{N_y}P(n_x a,n_y a,z)\ ,
\label{wall0}
\end{equation} 
represents the Polyakov loop averaged over the $xy$-plane at a given $z$.~\footnote{Here and in 
the following, $N_i$ ($i=x, y, z$) is the number of lattice sites in the $i$-direction.}
The wall average implies the projection at zero momentum in the $xy$-plane.

For the 2$^+$-channel, we used the variational 
method~\cite{Kronfeld,Luscher-Wolff} (for more details, see~\cite{FFGP07} and references therein.)

Our choice of wall-averaged operators in the 2$^+$-channel is
inspired by Ref.~\cite{Caselle1997} and reads 
\begin{equation}
\bar{P}_n(z)=\frac{1}{N_xN_y}\sum_{n_x=1}^{N_x} \sum_{n_y=1}^{N_y}P(n_x a,n_y a,z)
\biggl[P(n_x a+na,n_y a,z)-P(n_x a,n_y a+na,z)\biggr]\ .
\label{wall2}
\end{equation} 
In most cases we have taken 8 operators, corresponding to different values of $n$, with the 
largest $n$ almost reaching the spatial lattice size $N_x$. 

We consider also correlators of the (wall-averaged) real and imaginary parts of the Polyakov 
loop, defined as
\begin{eqnarray}
G_R(|z_1-z_2|) &=& \langle \mbox{Re} \bar P(z_1) \mbox{Re} \bar P(z_2)\rangle - 
\langle \mbox{Re} \bar P(z_1)\rangle \langle \mbox{Re} \bar P(z_2)\rangle \;,
\label{Dumitru2:1}\\
G_I(|z_1-z_2|) &=& \langle \mbox{Im} \bar P(z_1) \mbox{Im} \bar P(z_2)\rangle\;.
\label{Dumitru2:2}
\end{eqnarray}
The corresponding screening masses, $\hat m_R$ and $\hat m_I$, can be extracted in the 
same way as for the 0$^+$ mass.
We have studied the ratio $m_I/m_R$ over a wide interval of temperatures above the 
transition temperature $T_t$ of (3+1)$d$ SU(3) and seen how it compares with the prediction from 
high-temperature perturbation theory, according to which it should be equal to 
3/2~\cite{Nadkarni:1986cz,Dumitru:2002cf}, and with the prediction from the mean-field 
Polyakov loop model of Ref.~\cite{Pisarski0110214}, according to which it should be equal to 3 in 
the transition region. The interplay between the two regimes should delimit the 
region where mean-field Polyakov loop models should be effective.

\section{Numerical results}
\label{results}

We used the Wilson lattice action and generated Monte Carlo configurations by a combination 
of the modified Metropolis algorithm~\cite{Cabibbo-Marinari} with over-relaxation on
SU(2) subgroups~\cite{Adler}. The error analysis was performed by the jackknife method over bins 
at different blocking levels. We performed our simulations on a 16$^3\times$4 lattice, for which 
$\beta_t=5.6908(2)$~\cite{Boyd:1996bx}, over an interval of $\beta$ values ranging from
5.69 to 9.0. 

Screening masses are determined from the plateau of $m_{\mbox{\footnotesize eff}}(z)$ as a 
function of the wall separation $z$. In each case, the {\it plateau mass} is taken as the 
effective mass (with its error) belonging to the {\it plateau} and having the minimal 
uncertainty. We define {\it plateau} the largest set of consecutive data points, consistent with 
each other within 1$\sigma$.
This procedure is more conservative than identifying the plateau mass and its error
as the results of a fit with a constant on the effective masses $m_{\mbox{\footnotesize eff}}(z)$,
for large enough $z$. 

\begin{figure}[tb]
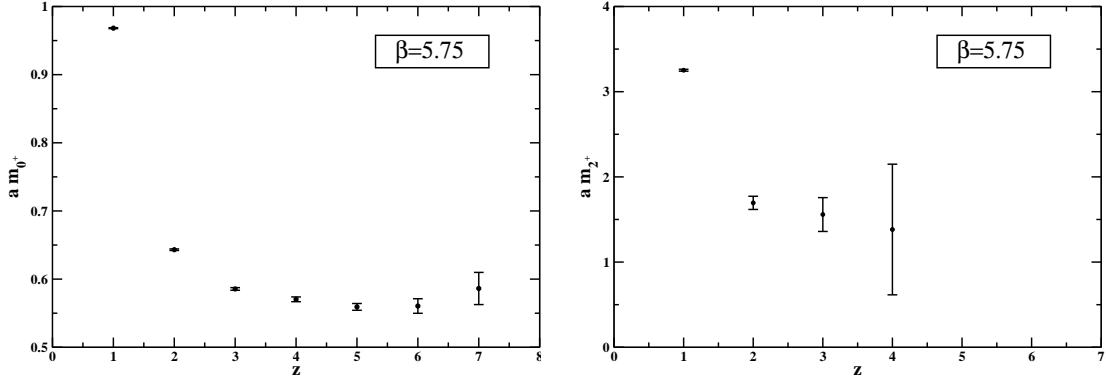

\centering
\bigskip
\includegraphics[scale=0.30]{massa0.eps} \quad 
\includegraphics[scale=0.30]{massa2.eps}
\caption[]{Effective mass in the 0$^+$ (left) and the 2$^+$ (right) channel  
as a function of the separation between walls on the $(x,y)$ plane at $\beta=5.75$.}
\label{eff_masses_5.75} 
\end{figure}

\begin{figure}[tb]
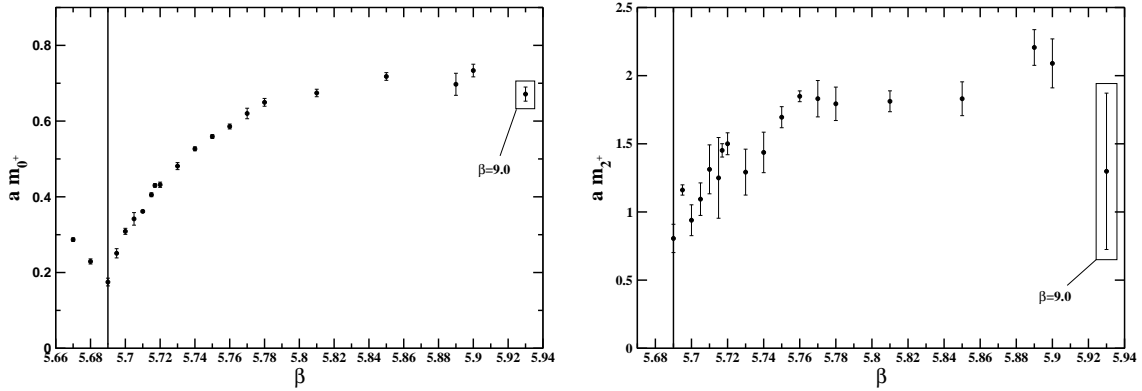

\centering
\bigskip
\includegraphics[scale=0.30]{mass0_vs_beta.eps} \quad
\includegraphics[scale=0.30]{mass2_vs_beta.eps}
\caption[]{Screening mass in the 0$^+$ channel (left) and in the 2$^+$ channel (right)
{\it vs.} $\beta$.}
\label{masses_vs_beta_1} 
\end{figure}

\begin{figure}[tb]
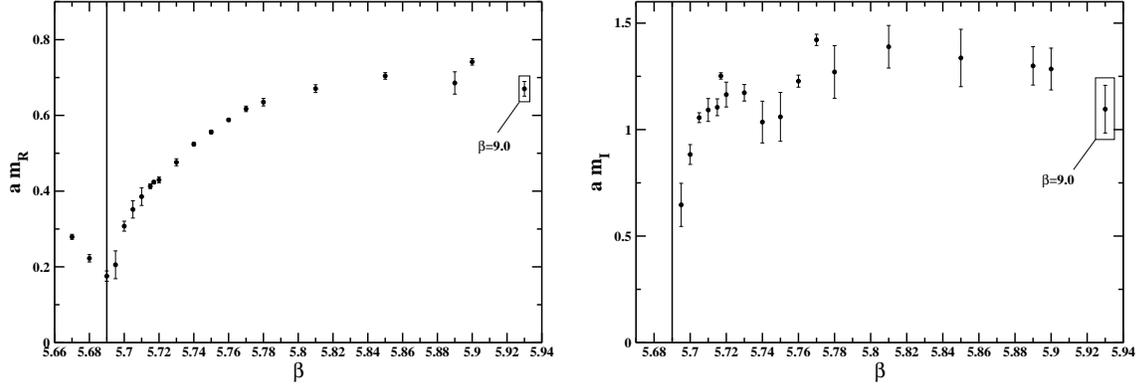

\centering
\bigskip
\includegraphics[scale=0.30]{mass_r_vs_beta.eps} \quad
\includegraphics[scale=0.30]{mass_i_vs_beta.eps}
\caption[]{Screening masses ${\hat m}_R$ (left) and ${\hat m}_I$ (right) {\it vs.} $\beta$.}
\label{masses_vs_beta_2} 
\end{figure}

Just above the critical value $\beta_t$ we find a large correlation length, which is not of 
physical relevance. It is instead a genuine finite size effect~\cite{Gavai-Karsch-Petersson} 
related to {\it tunneling} between degenerate vacua. This effect disappears by going to larger 
lattice volumes or moving away from $\beta_t$ in the deconfined phase. 
Tunneling can occur between the symmetric and the broken phase, and between
the three degenerate vacua of the deconfined phase. 
When tunneling is active,
the correlation function has the following expression~\cite{Gavai-Karsch-Petersson}:
\begin{equation}
G(|z_1-z_2|)\sim a_0e^{-m_0 |z_1-z_2|}+b_0 e^{-m_t |z_1-z_2|}\;,
\label{corr_funct_tunneling}
\end{equation}
where $m_t$ is the inverse of the tunneling correlation length and is generally much smaller 
than the fundamental mass $m_0$ and therefore behaves as a constant additive term
in the correlation function.~\footnote{In (\ref{corr_funct_tunneling}) 
we have taken into account only the 
lowest masses in the spectrum and, for brevity, omitted to write the ``echo'' terms.}
The dependence on $m_t$ in the correlation function can be removed by extracting the
effective mass by use of the combination
\begin{equation}
m_{\mbox{\footnotesize eff}}(z)= \ln \frac{G(z)-G(z+1)}{G(z+1)-G(z+2)} \ .
\label{m_eff_1}
\end{equation}
%%For $\beta \gtrsim 5.71$ the only active tunneling is among the three broken minima, and since
%%the separation among them is so clear, 
%%it is possible to ``rotate'' unambiguously all the configurations to the real sector 
%%and to treat them on the same ground.
A typical example of the behavior of the effective mass with $z$
is shown in Fig.~\ref{eff_masses_5.75} for the 0$^+$ and the 2$^+$ channels. 
In Figs.~\ref{masses_vs_beta_1} and~\ref{masses_vs_beta_2} we show the behavior with $\beta$ of
$\hat m_{0^+}$, $\hat m_{2^+}$, $\hat m_R$ and $\hat m_I$.
%%%%%%%%%%%%%%%%%%%%%%%%%%

\begin{figure}[tb]
\centering
\includegraphics[scale=0.30]{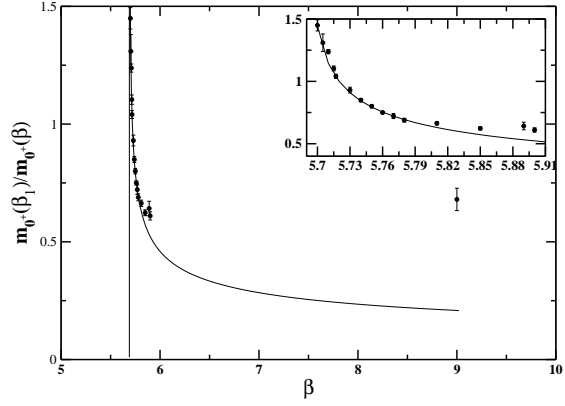}
%\centering
\caption[]{Comparison between the scaling function $[(\beta_1-\beta_t)/(\beta-\beta_t)]^{1/3}$
and the mass ratio $m_{0^+}(\beta_1)/m_{0^+}(\beta)$ for varying $\beta$, with $\beta_1$=5.72.}
\label{scaling} 
\end{figure}

We observe from that $\hat m_{0^+}$ and $\hat m_R$ are consistent within statistical errors, 
this indicating that the Polyakov loop correlation is dominated by the correlation between 
the real parts.
We can see that the fundamental mass in the 0$^+$ channel, as well as $\hat m_R$, 
becomes much smaller than 1 at $\beta_t$, as expected for a weakly 
first order phase transition.
In the cases of $\hat m_{0^+}$ and of $\hat m_R$ we have made some determinations 
{\it below} $\beta_t$ (see Figs.~\ref{masses_vs_beta_1} and \ref{masses_vs_beta_2}). 
It turns out that masses in lattice units take their minimum value just at $\beta_t$, where 
there is a ``cusp'' in the $\beta$-dependence. Such a behaviour was observed 
also by the authors of Ref.~\cite{Datta:2002je}, whose results, when the comparison is possible, 
agree with ours.
%%%%%%%%%%%%%%%%%%%%%%%%%%
%%%%%%%%%%%%%%%%%%%%%%%%%%
%%%%%%%%%%%%%%%%%%%%%%%%%%
%%%%%%%%%%%%%%%%%%%%%%%%%%
%%%%%%%%%%%%%%%%%%%%%%%%%%
%%\begin{figure}[tb]
%%\centering
%\includegraphics[width=5cm]{mass_i_vs_beta.eps} (a)
%\includegraphics[width=5cm]{mass_r_vs_beta.eps} (b)
%\includegraphics[width=11.5cm,bb=40 40 700 620]{mass_i_vs_beta.eps} (a)
%\includegraphics[width=11.5cm,bb=40 40 700 620]{mass_r_vs_beta.eps} (b)
%%\caption[]{Screening mass $\hat m_I$ (a) and $\hat m_R$ (b) as a function of $\beta$.}
%%\label{masses_ri_vs_beta} 
%%\end{figure}
%%%%%%%%%%%%%%%%%%%%%%%%%%
%\subsection{Scaling behavior and comparison with the Potts model}
We have also looked for a scaling law for the fundamental mass in the $0^+$ channel,
but with the understanding that any second-order-like scaling law, when applied to the 
region near a first order phase transition, should be taken as an {\it effective} description, 
which cannot hold too close to the transition point.
With this spirit, we have compared our data with the scaling law
\begin{equation}
\label{scal_rel}
\Bigg( \frac{\beta_1-\beta_t}{\beta_2-\beta_t} \Bigg)^{\nu} \sim 
\frac{\hat m_{0^+}(\beta_1)}{\hat m_{0^+}(\beta_2)} \ ,
\end{equation} 
where $\hat m_{0^+}(\beta_1)$ and $\hat m_{0^+}(\beta_2)$ are the 
fundamental masses in the $0^+$ channel at $\beta_1$ and $\beta_2$, respectively. 
We have considered several choices of $\beta_1$
and found that for each of them there is a wide ``window'' of $\beta$ values above $\beta_t$ where
the scaling law~(\ref{scal_rel}) works, with a ``dynamical'' exponent $\nu$ 
(see Ref.~\cite{FFGP07} for a details).
%Our results are summarized in Table.~\ref{scaling_fits}. 
For $\beta_1$=5.72 we have calculated also 
the $\chi^2$/d.o.f. when $\nu$ is put exactly equal to 1/3 (suggested in 
Ref.~\cite{Fisher-Berker} to apply 
to the {\it standard} correlation function), getting $\chi^2$/d.o.f.=0.75 in the window 
from $\beta=5.715$ to $\beta=5.78$. In Fig.~\ref{scaling} we show, for this choice of $\beta_1$, 
the comparison between data and the ``scaling'' function with $\nu$ set equal to 1/3.\\ 
%%%%%%%%%%%%%%%%%%%%%%%%%%
%\begin{minipage}{.45\textwidth}
%\end{minipage}
%\begin{minipage}[b]{.45\textwidth}
%\begin{table}[tb]
%\centering
%\caption{Summary of the fits of the mass ratios ${\hat m_{0^+}(\beta_1)}/{\hat m_{0^+}(\beta)}$ 
%with the function $((\beta_1-\beta_t)/(\beta-\beta_t))^\nu$. The second column
%contains the largest window of $\beta$ values for which the fit has a $\chi^2$/d.o.f. lower
%than 1.}
%\begin{tabular}{|l|c|l|l|}
%\hline
%$\beta_1$ & window of $\beta$ values & $\nu$ & $\chi^2$/d.o.f. \\
%\hline
%5.75  & 5.70  - 5.78 &  0.3619(90) & 0.90 \\
%5.74  & 5.70  - 5.78 &  0.365(11)  & 0.92 \\
%5.73  & 5.695 - 5.85 &  0.329(12)  & 0.77 \\
%5.72  & 5.695 - 5.78 &  0.354(12)  & 0.73 \\
%5.717 & 5.715 - 5.81 &  0.3228(94) & 0.69 \\
%5.715 & 5.695 - 5.78 &  0.358(10)  & 0.86 \\
%5.71  & 5.72  - 5.78 &  0.3951(76) & 0.33 \\
%5.705 & 5.705 - 5.78 &  0.448(22)  & 0.89 \\  
%5.70  & 5.695 - 5.90 &  0.3141(58) & 0.94 \\
%5.695 & 5.695 - 5.90 &  0.3095(67) & 0.41 \\
%\hline
%\end{tabular}
%\label{scaling_fits}
%\end{table}
%\end{minipage}
%%%%%%%%%%%%%%%%%%%%%%%%%%

\begin{figure}[tb]
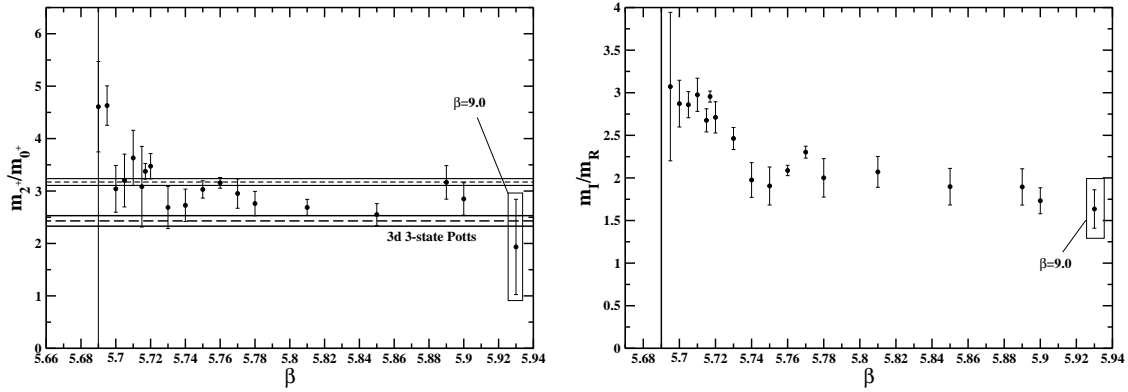

\centering
\bigskip
\includegraphics[scale=0.30]{ratio_20_new.eps}\quad
\includegraphics[scale=0.30]{ratio_ri.eps}
\caption[]{(Left) Ratio $m_{2^+}/m_{0^+}$ as a function of $\beta$ in the deconfined phase.
The three upper horizontal lines represent the constant (with its error) which fits the data (see 
the text for details); the three lower horizontal lines represent the corresponding mass ratio 
(with its error) found in the 3$d$ 3-state Potts model~\cite{FFGP06}. 
(Right) Ratio $m_I/m_R$ as a function of $\beta$ in the deconfined phase. The vertical 
line corresponds to the critical $\beta$ value.}
\label{ratio_20} 
\end{figure}

Then, we have considered the $\beta$-dependence of the ratio $m_{2^+}/m_{0^+}$, shown in
Fig.~\ref{ratio_20}. We have found that this ratio can be interpolated with a constant
in the interval from $\beta_t$ to $\beta=5.77$. This constant turned out to be
3.172(65), with a $\chi^2$/d.o.f equal to 1.085. In the fit we excluded the point at 
$\beta$=5.695, for which the determination of $m_{2^+}$ is probably to be rejected.  
If the point at $\beta$=5.695 is included, the constant becomes 3.214(64) with $\chi^2$/d.o.f 
=2.21.
The fact that the ratio $m_{2^+}/m_{0^+}$ is compatible with a constant in the mentioned interval
suggests that $\hat m_{2^+}$ scales similarly to $\hat m_{0^+}$ near the transition. This constant
turns out to be larger than the ratio between the lowest massive excitations in the
same channels in the broken phase of the 3$d$ 3-state Potts model, which was determined
in Ref.~\cite{FFGP06} to be 2.43(10).

We have calculated the ratio $m_I/m_R$ for $\beta$ ranging from 5.695 up to 9.0. We observe from 
the right panel of Fig.~\ref{ratio_20} that this ratio is compatible with 3/2 at the largest 
$\beta$ values considered, 
in agreement with the high-temperature perturbation theory. Then, when the temperature is lowered 
towards the transition, this ratio goes up to a value compatible with 3, in agreement with the 
Polyakov loop model of Ref.~\cite{Pisarski0110214}, which contains only quadratic, cubic and 
quartic powers of the Polyakov loop, i.e. the minimum number of terms required in order to be 
compatible with a first order phase transition. The same trend has been observed also in 
Ref.~\cite{Datta:2002je}.

%%%%%%%%%%%%%%%%%%%%%%%%%%
%%%%%%%%%%%%%%%%%%%%%%%%%%

\section{Conclusions and outlook}
In this work we have studied in the (3+1)$d$ SU(3) pure gauge theory above the deconfinement 
transition the lowest masses in the 0$^+$ and the 2$^+$ channels of angular momentum and parity 
and the screening masses resulting from the correlation between the real parts and between the 
imaginary parts of the Polyakov loop. 
The behavior of the ratio between the masses in the 0$^+$ and the 2$^+$ channels with the 
temperature suggests that they have a common scaling above the transition temperature. 
This ratio turns to be $\simeq$30\% larger than the ratio of the lowest massive excitations in 
the same channels of the 3$d$ 3-state Potts model in the broken phase. This can be taken
as an estimate of the level of approximation by which the Svetitsky-Yaffe conjecture, valid in 
strict sense only for continuous phase transitions, can play some role also for (3+1)$d$ SU(3) 
at finite temperature. 

The dependence on the temperature of the ratio between the screening masses from the
correlation between the real parts and between the imaginary parts of the Polyakov loop 
shows a nice interplay between the high-temperature regime, where perturbation theory should 
work, and the transition regime, where mean-field effective Polyakov loop models could 
apply.

\end{document}